\documentclass[twocolumn,showpacs,amsmath,amssymb,prb]{revtex4}

\usepackage{graphicx}
\usepackage[usenames]{color}

\newcommand{\up}{\uparrow}
\newcommand{\dn}{\downarrow}
\newcommand{\vareps}{\varepsilon}
\newcommand{\ket}[1]{\left|#1\right>}
\newcommand{\bra}[1]{\left<#1\right|}
\newcommand{\dt}[1]{\frac{\text{d}#1}{\text{d}t}}

\begin{document}

\title{Theory of transport through noncollinear single-electron spin-valve transistors}

\author{Stephan Lindebaum}
\affiliation{Theoretische Physik, Universit\"at Duisburg-Essen and CeNIDE, 47048 Duisburg}

\author{J\"urgen K\"onig}
\affiliation{Theoretische Physik, Universit\"at Duisburg-Essen and CeNIDE, 47048 Duisburg}

\date{\today}
\pacs{85.75.-d,73.23.Hk,85.35.Gv}

\begin{abstract}
We study the electronic transport through a noncollinear single-electron spin-valve transistor. It consists of a small metallic island weakly coupled to two ferromagnetic leads with noncollinear magnetization directions. The electric current is influenced by Coulomb charging and by spin accumulation. Furthermore, the interplay of Coulomb interaction and tunnel coupling to spin-polarized leads yields a many-body exchange field, in which the accumulated island spin precesses. We analyze the effects of this exchange field in both the linear and nonlinear transport regime. In particular, we find that the exchange field can give rise to a high sensitivity of the island's spin orientation on the gate voltage. 
\end{abstract}

\maketitle

\section{Introduction} \label{sec:introduction}
\begin{figure}[b]
		\includegraphics[width=.8\columnwidth]{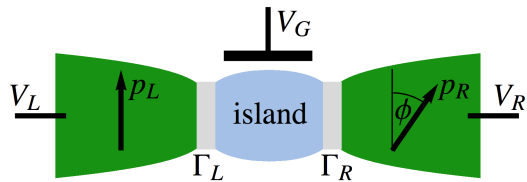}
		\caption{(Color online) Noncollinear single-electron spin-valve transistor: a metallic island is tunnel coupled to ferromagnetic source and drain leads, whose polarizations directions enclose an angle $\phi$.
		 }
		\label{fig:system}
	\end{figure}

The intense investigations in the field of spintronics have been partially driven by the prospect of developing new devices for current and future information technology.\cite{datta:1990,wolf:2001,greg:2002,zutic:2004} Continuing the trend of miniaturization down to the nanometer scale leads to devices in which Coulomb-interaction effects become important. Prominent examples are Coulomb blockade and Coulomb oscillations.\cite{devoret:1992} Hence, a fundamental understanding of the interplay between spin degrees of freedom and Coulomb interaction in nanoscale devices is required. A convenient model system to study this interplay is a single-electron spin-valve transistor, see Fig.~\ref{fig:system}. It consists of a metallic island that is tunnel coupled to ferromagnetic source and drain leads. The current flowing through the island is controlled by both an applied gate voltage and the relative orientation of the magnetization directions of source and drain lead. Experimental measurements in various versions of the single-electron spin-valve transistors with collinearly (parallel or antiparallel) polarized leads addressed current-voltage characteristics, tunnel magnetoresistance (TMR), spin accumulation, and magneto-Coulomb effects. This includes devices in which the central island is ferromagnetic (FFF),\cite{ootuka:1996,ono:1997,shimada:1998,ono:1998,brueckl:1998,takemura:2001,jedema:2002,shimada:2003,matsuda:2003,niizeki:2004,wunderlich:2006,seneor:2007} non-magnetic (FNF),\cite{seneor:2007,bernand:2006,liu:2007,bernand:2009} or even superconducting (FSF).\cite{chen:2002,johansson:2003,chen:2003,philip:2004,wang:2005} Many theoretical works also focused on collinear configurations of the leads' magnetization\cite{takahashi:1998,barnas:1998a,barnas:1998b,majumdar:1998,korotkov:1999,brataas:1999a,brataas:1999b,barnas:2000,brataas:2001,martinek:2002,weymann:2003,ernult:2007,barnas:2008} in order to discuss, e.g., an oscillating TMR with variation of bias voltage, an enhanced TMR in the Coulomb blockade regime, or a finite spin accumulation on the central island. But also noncollinear single-electron spin valve transistors have been discussed in the literature.\cite{brataas:2000,huertas:2000,braig:2005,wetzels:2005,wetzels:2006,linder:2007}
		
The single-electron spin-valve transistor is conceptually similar to a quantum-dot spin valve. In the latter, the central island hosting a continuum of single-particle energy levels is replaced by a quantum dot with a discrete level spectrum. This can be realized by either shrinking the central island in size or by using semiconductors or carbon nanotubes instead of metals. Quantum-dot spin valves have been studied extensively both theoretically\cite{bulka:2000,koenig:2003,martinek:2003,martinek:2003a,cottet:2004,cottet:2004a,braun:2004,weymann:2005,weymann:2005a,fransson:2005,fransson:2005a,rudzinski:2005,weymann:2006,braun:2006,simon:2007,splettstoesser:2008,lindebaum:2009,sothmann:2010,sothmann:2010a,baumgaertel:2010} and experimentally.\cite{yakushiji:2001,deshmukh:2002,pasupathy:2004,zhang:2005,sahoo:2005,hamaya:2007,hauptmann:2008,merchant:2008,hofstetter:2010} One intriguing prediction\cite{koenig:2003,braun:2004,martinek:2003} for the quantum-dot spin valve was the existence of an interaction-induced exchange field that acts on the spins of the quantum-dot electrons as a consequence of the tunnel coupling to spin-polarized leads. This exchange field, which is tunable by gate and bias voltage, leads to a precession of an accumulated dot spin\cite{koenig:2003,braun:2004} or a splitting in the Kondo resonance.\cite{martinek:2003} The latter has been experimentally confirmed recently.\cite{pasupathy:2004,hamaya:2007,hauptmann:2008}
	
The origin of the exchange field is a level renormalization that is spin dependent as a consequence of spin-dependent tunnel couplings. This idea has later been transferred to describe gate-dependent tunneling-induced level shifts in carbon nanotubes with orbital-dependent tunnel couplings to normal leads\cite{holm:2008} and to molecular systems with different tunnel couplings of degenerate molecular states to two normal leads.\cite{darau:2009} It is quite natural to expect the existence of a similar exchange field for a single-electron spin-valve transistor. The implications of such an exchange field on the linear conductance has already been discussed in the Coulomb-blockade regime.\cite{wetzels:2005,wetzels:2006} In the present paper, we consider the same system and derive kinetic equations for the island charge and spin within a diagrammatic real-time transport formalism up to the lowest order in the tunnel-coupling strength (sequential-tunneling limit). The resulting kinetic equations, which agree with Refs.~\onlinecite{wetzels:2005,wetzels:2006} in the linear-response regime, are the basis for the discussion of the spin accumulation and its implication on transport in linear {\it and} nonlinear response. The used formalism systematically includes all contributions to a given order in the perturbation expansion. In addition to the spin-dependent tunneling rates, the exchange field automatically emerges in the kinetic equations as a result of the considered Hamiltonian.

\section{Model}	\label{sec:model}
	The single-electron spin-valve transistor shown in Fig.~\ref{fig:system} is modeled by the total Hamiltonian $H = H_I + H_C+H_L+H_R +H_T$. The first part,
	\begin{equation}\label{eq:hamiltonianI}
		H_I=\sum_{l\sigma\nu}\vareps_{l}\,c_{l\sigma\nu}^\dagger c_{l\sigma\nu}\,,
	\end{equation}
	describes the metallic island whose energy spectrum $\vareps_{l}$ is characterized by a small level spacing $\Delta\vareps$. The index $l$ labels the energy levels of the island, $\sigma\in\{\up,\dn\}$ the spin, and $\nu=1,\ldots, N_c$ is the transverse channel index. The annihilation (creation) operator of island electrons in the state $l\sigma\nu$ is denoted by $c_{l\sigma\nu}^{(\dagger)}$. We assume that the energy levels are independent of spin and transverse channel number. We will consider the limit of temperature and bias voltage larger than the level spacing, $k_BT,eV\gg \Delta\vareps$, for which the energy spectrum can be viewed as continuous.
  	
	The Coulomb interaction of the island electrons is accounted for by the charging-energy term
	\begin{equation}\label{eq:hamiltonianC}
	H_C=E_C(N-N_\text{ext})^2\,,
	\end{equation}
	where $N$ is the number of electrons on the island. The charging energy scale $E_C=e^2/(2C_\Sigma)$ is determined by the total capacitance $C_\Sigma$, which is the sum of the capacitances of the two tunnel junctions, $C_L,C_R$, and the gate, $C_G$. For equal capacitances of the two tunnel junctions and a symmetrically applied transport voltage, the external charge $e\,N_\text{ext}=C_GV_G$ depends on the gate voltage $V_G$ only. For later convenience, we define $\Delta_N$ as the difference of charging energies of $N+1$ and $N$ electrons, i.e., $\Delta_N=E_C [2(N-N_\text{ext})+1]$. 

Each of the leads is described as a reservoir of noninteracting fermions:
	\begin{equation}\label{eq:hamiltonianLR}
		H_r = \sum_{k{s}\nu} \epsilon_{rk{s}}^{}\, a_{rk{s}\nu}^\dagger a_{rk{s}\nu}\,,
	\end{equation}
	with indices for lead $r \in \{L,R\}$, momentum $k$ and $a_{rk{s}\nu}^{(\dagger)}$ is the annihilation (creation) operator of lead $r$. The index ${s}=+(-)$ denotes the majority (minority) spin states (along the magnetization direction ${\bf\hat{n}}_r$ of the respective lead) with the density of states $\rho_{s}^{r}$, which we assume to be energy independent. The lead's degree of spin polarization (at the Fermi energy) is characterized by  $p_r= (\rho_{+}^{r}-\rho_{-}^{r})/(\rho_{+}^{r}+\rho_{-}^{r})$. The angle enclosed by ${\bf\hat{n}}_L$ and ${\bf\hat{n}}_R$ is denoted by $\phi$.

\begin{figure}
		\centering
		\includegraphics[width=.8\columnwidth]{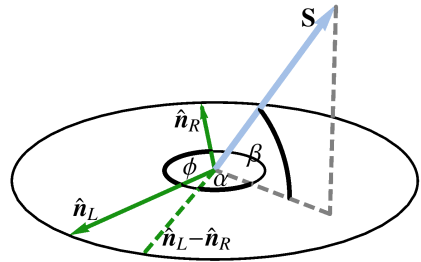}
		\caption{(Color online) Scheme of relation between defined angles $\alpha,\beta$, the polarization directions of the two leads ${\bf\hat{n}}_L,{\bf\hat{n}}_R$, and the accumulated spin on the island ${\bf S}$.}
		\label{fig:angles}
	\end{figure}

The tunneling Hamiltonian $H_T = \sum_r H_{T,r}$\;, with
	\begin{equation}\label{eq:hamiltonianT}
		H_{T,r}=\sum_{kl{s}\sigma\nu}\,V^{r}_{k{s}\sigma\nu}\,a^\dag_{rk{s}\nu} c_{l\sigma\nu} + \text{H.c.}\, ,
	\end{equation}
describes tunneling between island and leads. Both the spin and the transverse channel index $\nu$ are conserved during tunneling. The latter is obvious from the fact that the tunneling Hamiltonian is diagonal in $\nu$. In the following, we assume the tunneling matrix elements $V^{r}_{k{s}\sigma\nu}=V^{r}_{{s}\sigma}$ to be independent of momentum $k$ and transverse channel index $\nu$. Spin conservation is accounted for by expressing $V^{r}_{{s}\sigma}$ as a product of the (spin-independent) tunnel amplitudes $t_r$ and the matrix elements of an SU(2) rotation that connects the (in general) different spin quantization axes for the two leads and the island. For our derivation of the kinetic equations it is convenient to choose the spin quantization axis of the island ${\bf\hat{n}}_S$ along the direction of the accumulated island spin ${\bf S}$. Its orientation relative to the lead's magnetization directions can be parametrized by two angles, as shown in Fig.~\ref{fig:angles}: the angle $\alpha$ enclosed by the ${\bf\hat{n}}_L\!\!-\!{\bf\hat{n}}_R$-axis and the projection of ${\bf S}$ onto the $({\bf\hat{n}}_L,{\bf\hat{n}}_R)$-plane, and the angle $\beta$ between ${\bf S}$ and the $({\bf\hat{n}}_L,{\bf\hat{n}}_R)$-plane itself. 
Then, the tunnel matrix elements $V^{r}_{s\sigma}$ for the left lead become
 	\begin{eqnarray}\label{eq:tme}
		V^L_{\pm\up}=t_L\frac{\pm e^{i\phi /2}\cos\left(\frac{\beta}{2}-\frac{\pi}{4}\right)-i e^{i\alpha}\sin\left(\frac{\beta}{2}-\frac{\pi}{4}\right)}{\sqrt{2}},\\
		V^L_{\pm\dn}=t_L\frac{\pm e^{i\phi /2}\sin\left(\frac{\beta}{2}-\frac{\pi}{4}\right)+i e^{i\alpha}\cos\left(\frac{\beta}{2}-\frac{\pi}{4}\right)}{\sqrt{2}},
	\end{eqnarray}
 	while the elements of the right lead are described by the same expressions but with the replacements $L\rightarrow R$ and $\phi\rightarrow -\phi$. The tunneling rate for electrons from lead $r$ with spin $s$ into the island spin state $\sigma$ is quantified by $ \Gamma_{s\sigma}^r/\hbar =  2 \pi \rho_{s}^{r} |V^{r}_{s\sigma}|^2/\hbar $. In addition, we define $\Gamma_\sigma^r=\sum_s\Gamma_{s\sigma}^r$, $\Gamma_r=\sum_\sigma\Gamma_\sigma^r/2$, as well as $\Gamma=\sum_r \Gamma_r$.

\section{Method}	\label{sec:method}
	The dynamics of a quantum-mechanical system is determined by the time evolution of its total density matrix. Since we do not want to investigate the behavior of the lead degrees of freedom, we integrate them out and obtain a reduced density matrix $\hat{\rho}_\text{red}$, which only contains the degrees of freedom of the island. An island state is characterized by the ket vector $\ket{\chi}=\ket{ \{ n_{l\sigma \nu} \}}$, where $n_{l\sigma \nu}	=0,1$ counts the occupation of the corresponding island level. In this notation, the elements of the reduced density matrix are defined as $P^{\chi_1}_{\chi_2}=\bra{\chi_1}\hat{\rho}_\text{red}\ket{\chi_2}$. Diagonal elements $P^{\chi}_{\chi}\equiv P_{\chi}$ describe the probabilities to find the island in the corresponding state $\ket{\chi}$. For later convenience, we introduce the notation $\ket{\chi}=\ket{(\bar \chi_{l\nu};\psi)}$, where $\bar \chi_{l\nu}$ describes the occupation of all island levels except for the orbital $l$ and channel index $\nu$, while $\psi \in \{0,\up,\dn,\text{d}\}$ explicitly shows the occupation of level $l$ and channel $\nu$. 
	
	The time evolution of $\hat{\rho}_\text{red}$ is determined by the following kinetic equations:
	\begin{eqnarray}\label{eq:mastereq}
		\nonumber\dt{}{P_{\chi_2}^{\chi_1}}(t)=&&\hspace{-.25cm}-\frac{i}{\hbar}\left(E_{\chi_1}-E_{\chi_2}\right)P_{\chi_2}^{\chi_1}(t)\\
		&&\hspace{-.25cm}+\int_{-\infty}^t\text{d}t'\sum_{\chi'_1\chi'_2}W_{\chi_2\,\chi'_2}^{\chi_1\,\chi'_1}(t,t'\,)P_{\chi'_2}^{\chi'_1}(t'\,)\;.
	\end{eqnarray}
	Here, the definition of the energies of the reduced system, $E_\chi=\bra{\chi}(H_I+H_C)\ket{\chi}$, is used. The kernels $W_{\chi_2\,\chi'_2}^{\chi_1\,\chi'_1}(t,t'\,)$ describe transitions between the matrix elements  $P_{\chi'_2}^{\chi'_1}(t')$ and $P_{\chi_2}^{\chi_1}(t)$ of the initial and final density matrix, respectively. They can be obtained using a diagrammatic real-time technique formulated on the Keldysh contour. It allows for a systematic perturbative expansion in the coupling strength. In this paper, we truncate the expansion at the lowest order $\Gamma$ to describe the weak-coupling limit (sequential tunneling). In general, Eq.~(\ref{eq:mastereq}) describes non-Markovian behavior, i.~e., the time evolution of matrix element $P_{\chi_2}^{\chi_1}(t)$ depends on the reduced density matrix at previous times $t'$. In the stationary limit, however, all matrix elements of $\rho_\text{red}$ become time independent ($P_{\chi_2}^{\chi_1}(t)=P_{\chi_2}^{\chi_1}(t')=P_{\chi_2}^{\chi_1}$), and with the additional definition $W_{\chi_2\,\chi'_2}^{\chi_1\, \chi'_1}=\int_{-\infty}^t\text{d}t'\;W_{\chi_2\,\chi'_2}^{\chi_1\,\chi'_1}(t,t'\,)$, Eq.~(\ref{eq:mastereq}) simplifies to 
	\begin{eqnarray}\label{eq:meqsimp}
		0=\dt{}{P_{\chi_2}^{\chi_1}}=-\frac{i}{\hbar}\left(E_{\chi_1}-E_{\chi_2}\right)P_{\chi_2}^{\chi_1} +\sum_{\chi'_1,\chi'_2}W_{\chi_2\, \chi'_2}^{\chi_1\,\chi'_1}P_{\chi'_2}^{\chi'_1}\;.
	\end{eqnarray}
The stationary charge current through lead $r$ is given by
	\begin{equation}\label{eq:current}
		I_r=\sum_{\chi,\chi'_1,\chi'_2}\left(W^{I_r}\right)_{\chi\, \chi'_2}^{\chi\, \chi'_1}P_{\chi'_2}^{\chi'_1} \, ,
	\end{equation}
	where the matrix elements of the current transition rates $W^{I_r}$ are directly obtained by multiplying the corresponding matrix elements of $W$ with the net transported charge from lead $r$ to the island. As we consider the stationary state, the charge on the island is conserved. Hence we are able to define the current flowing through the device as $I=I_L=-I_R$. For a detailed derivation of this diagrammatic language and its rules for the calculation of diagrams, we refer to Refs.~\onlinecite{diagrams1,diagrams2,technique1,technique2,braun:2004}.
	
	For a metallic single-electron transistor with a dense level spectrum on the island, the dimension of the reduced density matrix and, thus, the number of kinetic equations determining them is huge. A drastic simplification of the problem is achieved by getting rid of the off-diagonal matrix elements of the reduced density matrix on the right-hand side of the kinetic equations. Since the tunneling Hamiltonian conserves charge and is diagonal in the transverse channel index, only those matrix elements $P_{\chi'}^{\chi}$ of states $|\chi\rangle=|\{n_{l\sigma\nu}\}\rangle$ and $|\chi'\rangle=|\{n'_{l\sigma\nu}\}\rangle$ with the same number of island electrons in any channel $\nu$ need to be considered, $\sum_{l\sigma} n_{l\sigma\nu}= \sum_{l\sigma} n'_{l\sigma\nu}$. A further simplification relies on the assumption that there is a fast, spin-independent energy relaxation within the island, i.e., the time scale of energy-relaxation $\tau_\text{er}$ is smaller than the dwell time $\tau_\text{dw}$ (which, in turn, is smaller than the intrinsic spin-flip time $\tau_\text{sf}$ in order to sustain spin imbalance on the island). This is reasonable due to the fact that in contrast to quantum dots, metallic islands accommodate a large number of electrons and hence exhibit many relaxation channels. As a consequence, any coherent superposition between states with different occupations of the orbital levels $l$ are destroyed and only coherences between different spin states is kept. This leads to the more restrictive condition $\sum_{\sigma} n_{l\sigma\nu}= \sum_{\sigma} n'_{l\sigma\nu}$. Since the spectrum on the island is spin degenerate, energy differences $E_\chi-E_{\chi'}$, that would appear on the right-hand side of the kinetic equation, vanish. 

	To remove the remaining coherent superpositions of different spin states we choose the spin quantization axis on the island along the direction of the accumulated spin and, furthermore, assume that the steady-state spin structure of the island is rotationally invariant about the quantization axis. The latter assumption neglects any anisotropies of quadrupole and higher moments within the plane perpendicular to the dipole moment. Eventually, we conclude $n_{l\sigma\nu}=  n'_{l\sigma\nu}$ for the proper choice of the spin quantization axis, i.e., only diagonal matrix elements $P_\chi \equiv P_\chi^\chi$ enter the right-hand side of the kinetic equations in the stationary limit,
	\begin{equation}\label{eq:simpleME}
  		0=  \dt{}{P_{\chi_2}^{\chi_1}}= \sum_{\chi} W_{\chi_2 \, \chi}^{\chi_1 \, \chi} P_{\chi} \, ,
	\end{equation}
	and similar for the current.

The assumed fast, spin-independent energy relaxation within the island does not only destroy coherences of states with different occupations of the orbital levels $l$, it also leads to a thermal equilibrium among all states that are connected by this relaxation, i.e., states with given numbers $N_\up$ and $N_\dn$ of spin $\uparrow$ and $\downarrow$ electrons on the island. The individual occupation for level $l$ with spin $\sigma$ and channel index $\nu$ under the condition that the island contains $N_\up$ and $N_\dn$ electrons with spin $\up$ and $\dn$, respectively, can be expressed by the conditional probabilities $F(l\sigma\nu |N_\up,N_\dn)$. Beenakker first introduced these probability functions to discuss resonant tunneling through a quantum dot coupled to two electron reservoirs,\cite{FBeenakker} while Barna\'s \textit{et al.} used them in the context of collinear single-electron spin-valve transistors.\cite{collfmSET} In thermal equilibrium, $F(l\sigma\nu |N_\up,N_\dn)$ reduces to the Fermi function with a spin-dependent chemical potential, $f(\varepsilon_l-\mu_\sigma(N_\sigma))$ with $f(E)=1/[\exp(\frac{E}{k_BT})+1]$. The spin-dependent chemical potential $\mu_\sigma(N_\sigma)$ is determined by the condition 
	\begin{equation}
		N_\sigma = \sum_l f\left[\epsilon_l-\mu_\sigma(N_\sigma)\right] \, . 
	\end{equation}
 
	The kernels $W_{\chi_2 \, \chi}^{\chi_1 \, \chi}$ in Eq. (\ref{eq:simpleME}) depend on the initial state $\chi$. But most of the information about the individual occupations of the levels contained in $\chi$ are irrelevant for the evaluation of the kernel matrix element. What matters is the occupation and the energy of only those island levels that are involved in the tunneling processes that take place in the transition described by the considered kernel. In this paper, we restrict ourselves to first order in the tunnel-coupling strength. Each kernel is, then, a sum over contributions for which only one level $l$ and transverse channel index $\nu$ is involved (we remind that coherent superpositions of different levels $l$ or transverse channels $\nu$ do not appear). In the notation $\ket{\chi}=\ket{(\bar \chi_{l\nu};\psi)}$, which makes the occupation of level $l$ and channel index $\nu$ explicit, the non-vanishing kernel matrix elements can be written in the form $W_{(\bar \chi_{l\nu};\psi_2) \, (\bar \chi_{l\nu};\psi )}^{(\bar \chi_{l\nu};\psi_1 ) \, (\bar \chi_{l\nu};\psi )}$, i.e., the part $\bar \chi_{l\nu}$ is not changed during the transition. The value of the kernel depends on the total island charge $N \equiv N_\up + N_\dn$ via the charging-energy contribution to the Hamiltonian and the occupation and energy of only the selected level $l$ and transverse channel index $\nu$. This occupation, however, is fully determined by the spin-dependent chemical potential $\mu_\sigma(N_\sigma)$, which, in turn, depends on $N_\up$ and $N_\dn$.
 
	For an island with a dense level spectrum, the dependence of the chemical potential $\mu_\sigma(N_\sigma)$ on $N_\sigma$ is rather weak: the change for each added electron with spin $\sigma$ is of the order of the mean level spacing $\Delta \vareps$ and, thus, much smaller than the energy scales $k_BT,eV$ relevant for transport. Therefore, we use $\mu_\sigma(N_\sigma) \approx \mu_\sigma$ independent of $N_\sigma$ in the following. This has the consequence that the value of the kernel matrix element $W_{\chi_2 \, \chi}^{\chi_1 \, \chi}$ depends only on the total island charge $N$ (of state $\chi$, i.e., $N=\sum_\sigma N_\chi^\sigma$ with $N_\chi^\sigma=\sum_{l\nu } \langle \chi |c^\dagger_{l\sigma \nu} c_{l\sigma \nu } | \chi \rangle$), the energy and occupation of the island level $l$ with channel index $\nu$ that is involved in the transition. This dependence can be cast in the notation $W_{\psi_2 \, \psi}^{\psi_1 \, \psi} (l,\nu,N)$, where the occupation of the levels and channels different from $l$ or $\nu$ does not appear explicitly. 
	\begin{figure}
		\includegraphics[width=.5\columnwidth,angle=270]{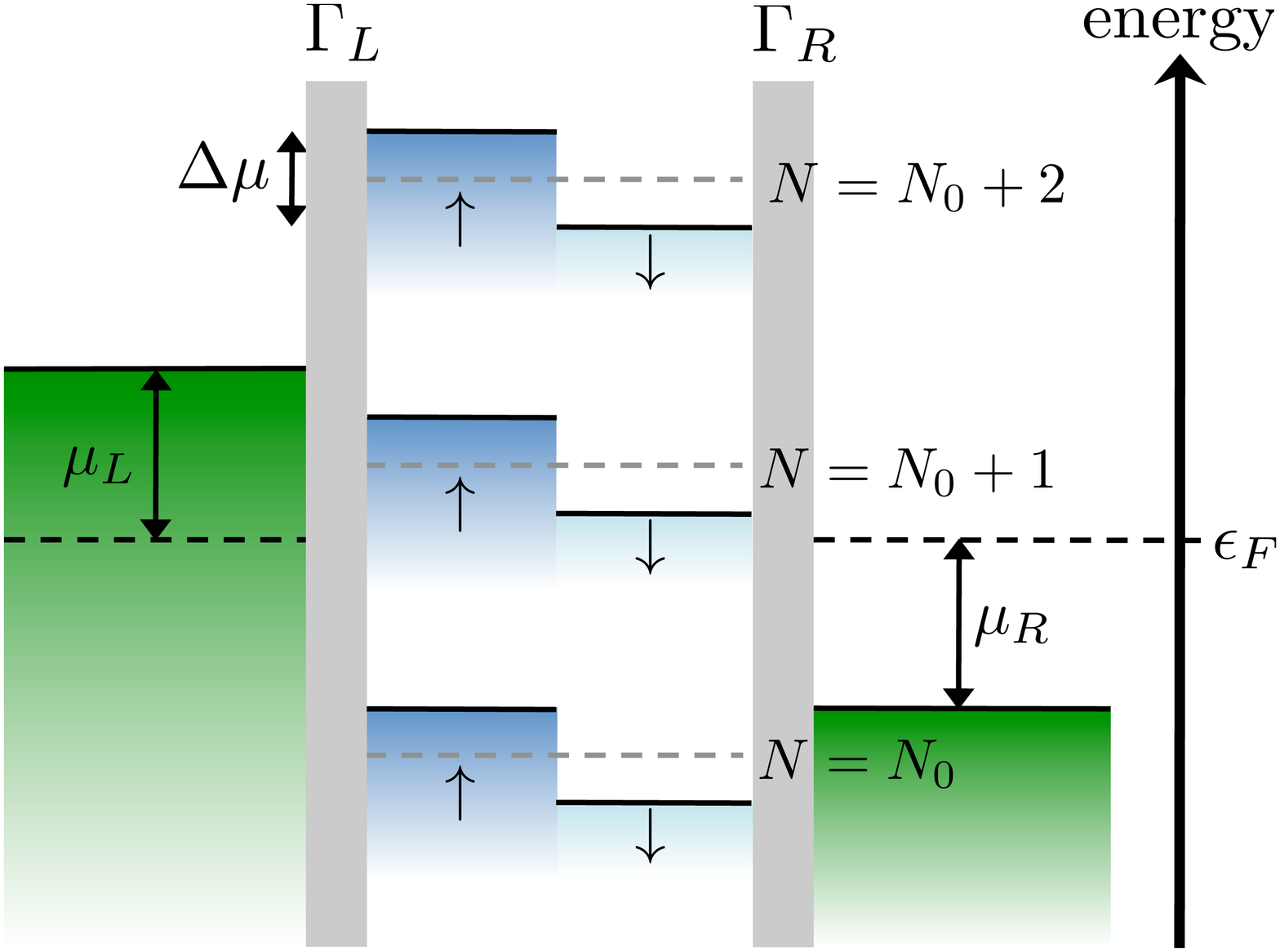}
		\caption{(Color online) Energy scheme of single-electron spin-valve transistor. The island is described as two independent electron reservoirs related to the spin species $\sigma\in\{\up,\dn\}$. The chemical potentials $\mu_{L,R}$ describe the applied bias voltage and $\Delta\mu=\mu_\up-\mu_\dn$ determines the spin accumulation on the central electrode. The charge states $N$ are tunable via gate voltage and $\epsilon_F$ denotes the Fermi energy.}
		\label{fig:energy}
	\end{figure}
	
	The independent degrees of freedom governed by the kinetic equations are the probabilities to find $N$ electrons on the island,
	\begin{equation}
			P_N = \sum_\chi P_\chi \delta_{N,N_\chi} \, ,  
	\end{equation}
	as well as the three components of the total island spin, or, equivalently, the magnitude and the direction of the island spin. The latter enter the right-hand side of the kinetic equations through the spin splitting $\Delta \mu \equiv \mu_\up-\mu_\dn$ of the chemical potential, $S= \hbar \rho_I \Delta\mu/2$, and the angles $\alpha$ and $\beta$ appearing in the tunnel matrix elements. We remark that, for a constant density of states on the island, the average chemical potential $\mu \equiv (\mu_\up+\mu_\dn)/2$ does not depend on the amplitude of the accumulated spin $S$. For each independent degree of freedom we need one kinetic equation. These are provided by the kinetic equations for the following operators $\hat A$: the projector $\ket{N}\bra{N} = \sum_\chi |\chi\rangle \langle \chi | \delta_{N,N_\chi}$ on the state with total island charge $N$ and the total island spin ${\bf \hat{S}} = (\hbar/2) \sum_{l\sigma \sigma' \nu} c_{l\sigma\nu}^\dagger {\vec{\sigma}}_{\sigma \, \sigma'} c_{l\sigma'\nu}$, where $\vec{\sigma}$ is the vector of Pauli spin matrices. The kinetic equations are given by
	\begin{eqnarray}
		0=\dt{}P_N &=&\sum_{\chi, \chi' } \delta_{N,N_\chi} W_{\chi \, \chi'}^{\chi \, \chi'} P_{\chi'}\\
		0=\dt{}{\langle {\bf \hat{S}} \rangle}&=&\sum_{\chi_1 , \chi_2 , \chi'} \bra{\chi_2} {\bf \hat{S}} \ket{\chi_1} W_{\chi_2 \, \chi'}^{\chi_1 \, \chi'} P_{\chi'} \, .
	\end{eqnarray}
	Introducing all the simplifications on the right-hand side, as discussed above, and making use of the notation $f^+_\sigma(\epsilon_l) = f(\epsilon_l - \mu_\sigma)$ and $f^-_\sigma(\epsilon_l) = 1-f(\epsilon_l - \mu_\sigma)$ leads to
	\begin{eqnarray}
		0=\dt{}P_N &=& \sum_{l\nu} W_{0 \, 0}^{0 \, 0} (l,\nu,N) f^-_\up (\epsilon_l) f^-_\dn (\epsilon_l ) P_N \nonumber \\
	 	&+& \sum_{l\nu\sigma}  W_{\sigma \, 0}^{\sigma \, 0} (l,\nu,N-1) f^-_\up(\epsilon_l )f^-_\dn(\epsilon_l ) P_{N-1} \nonumber \\
		&+& \sum_{l\nu\sigma}  W_{0 \, \sigma}^{0 \, \sigma} (l,\nu,N+1) f^+_\sigma(\epsilon_l)f^-_{\bar\sigma}(\epsilon_l ) P_{N+1} \nonumber \\
		&+& \sum_{l\nu\sigma} W_{\sigma \, \sigma}^{\sigma \, \sigma} (l,\nu,N) f^+_\sigma(\epsilon_l )f^-_{\bar\sigma} (\epsilon_l ) P_{N}\nonumber \\
		&+& \sum_{l\nu\sigma}  W_{d \, \sigma}^{d \, \sigma} (l,\nu,N-1) f^+_\sigma(\epsilon_l )f^-_{\bar\sigma}(\epsilon_l ) P_{N-1}\nonumber \\
		&+& \sum_{l\nu\sigma}  W_{\sigma \, d}^{\sigma \, d} (l,\nu,N+1) f^+_\up(\epsilon_l )f^+_\dn(\epsilon_l ) P_{N+1} \nonumber \\
		&+& \sum_{l\nu}  W_{d \, d}^{d \, d} (l,\nu,N) f^+_\up(\epsilon_l )f^+_\dn (\epsilon_l) P_{N}\, ,
	\end{eqnarray} 
for the probabilities $P_N$. For the $z$ component of the island spin we obtain
	\begin{eqnarray}\label{eq:MESz}
		0\!=\!\dt{}\langle S_z \rangle\!\!\!&=&\!\!\!\hbar\!\!\sum_{Nl\nu\sigma} \!\!\sigma W_{\sigma \, 0}^{\sigma \, 0} (l,\nu,N) f^-_\up(\epsilon_l )f^-_\dn(\epsilon_l ) P_{N}\nonumber \\
		&+&\!\!\!\hbar\!\!\sum_{Nl\nu\sigma} \!\!\sigma W_{\sigma \, \sigma}^{\sigma \, \sigma} (l,\nu,N) f^+_\sigma(\epsilon_l )f^-_{\bar\sigma} (\epsilon_l ) P_{N}\nonumber \\
		&+&\!\!\!\hbar\!\!\sum_{Nl\nu\sigma}  \!\!\sigma W_{\sigma \, d}^{\sigma \, d} (l,\nu,N) f^+_\up(\epsilon_l )f^+_\dn(\epsilon_l ) P_{N},
	\end{eqnarray} 
	where $\sigma$ contributes to sums with a $+$ sign for $\sigma=\up$ and a $-$ sign for $\sigma=\dn$. The kinetic equations for the $x-$ and $y$-components of the island spin, expressed by the raising and lowering operators $\langle {\bf S}^\pm \rangle = \langle {\bf S}_x \pm i {\bf S}_y \rangle$, are
	\begin{eqnarray}\label{eq:MESp}
		0\!=\!\dt{}\langle {\bf S}^+ \rangle\!\!\!&=&\!\!\!\hbar\sum_{Nl\nu} W_{\up \, 0}^{\dn \, 0} (l,\nu,N) f^-_\up(\epsilon_l )f^-_\dn(\epsilon_l ) P_{N}\nonumber \\
		&+&\!\!\!\hbar\!\sum_{Nl\nu\sigma} W_{\up \, \sigma}^{\dn \, \sigma} (l,\nu,N) f^+_\sigma(\epsilon_l )f^-_{\bar\sigma} (\epsilon_l ) P_{N}\nonumber \\
		&+&\!\!\!\hbar\!\sum_{Nl\nu} W_{\up \, d}^{\dn \, d} (l,\nu,N) f^+_\up(\epsilon_l )f^+_\dn(\epsilon_l ) P_{N},
	\end{eqnarray} 
	and the kinetic equation for $\langle {\bf S}^- \rangle$ is the same but replacing $W_{\up \, \psi}^{\dn \, \psi} (l,\nu,N)$ with $W_{\dn \, \psi}^{\up \, \psi} (l,\nu,N)$ on the right-hand side.

	The explicit values of $W_{\psi_2 \, \psi}^{\psi_1 \, \psi} (l,\nu,N)$ are given in Appendix \ref{ap:kerele}. We plug them in and replace the summation over the island levels $l$ by an integration over an energy $\omega$, multiplied by the density of states of the given spin $\rho_I^\sigma$. In the present paper, we assume the central electrode to be nonmagnetic, hence the island density of states is spin independent, $\rho_I^\sigma=\rho_I/2$. The occurring integrals can be carried out and we finally get the following master equations:
	\begin{eqnarray}\label{eq:MEdiagonal}	
		\nonumber\dt{}P_{N}&=&\pi\sum_{r\sigma}\left[\alpha_{r\sigma}^{+}(\Delta_{N-1})P_{N-1}+\alpha_{r\sigma}^{-}(\Delta_{N} )P_{N+1}\right.\\
		\label{eq:mefinalP*}&&\qquad\left.-\alpha_{r\sigma}^{+}(\Delta_{N} )P_{N}-\alpha_{r\sigma}^{-}(\Delta_{N-1} )P_{N}\right],\\
		\nonumber\dt{}\langle S_z \rangle&=&\pi\hbar \sum_{Nr\sigma}\sigma \left[\alpha_{r\sigma}^{+}(\Delta_N)-\alpha_{r\sigma}^{-}(\Delta_{N-1} )\right] P_{N}.
	\end{eqnarray}
	Here, we used the definition of the island rate functions
	\begin{eqnarray}
		\alpha_{r\sigma}^{\pm}(E):=\pm\alpha_{r\sigma}^0\;\frac{E-(\mu_r-\mu_\sigma)}{\exp\left[\pm\frac{E-(\mu_r-\mu_\sigma)}{k_B T}\right]-1},
	\end{eqnarray}
	where $\alpha_{r\sigma}^0=\frac{\rho_IN_c}{2\pi\hbar}\Gamma_\sigma^r$ is the dimensionless conductance of lead $r$ for island spin $\sigma$, $N_c$ the number of transverse channels, and $E$ is the energy of the tunneling electron. The equations (\ref{eq:mefinalP*}) are linearly dependent. Hence, to solve the master equations we need an additional equation that is provided by the normalization condition $\sum_{N}P_{N}=1$. For collinear single-electron spin-valve transistors $(\phi\in\{0,\pi\})$, the spin accumulation naturally points in a direction parallel to the lead magnetization directions $\hat{{\bf n}}_r$, this directly defines the angles $\alpha$ and $\beta$. Therefore, the equations in Eq. (\ref{eq:MEdiagonal}) are sufficient to describe such a system but not the noncollinear case.
		
	In the derivation of Eqs. (\ref{eq:MESz})-(\ref{eq:MEdiagonal}), we have used a specific coordinate system for the spin. In the chosen coordinate system, the $x$ and $y$ components of ${\bf S}$ vanish. The final version of the kinetic equations for the spin can, however, be written in a coordinate-free representation. First, we proceed in a manner analogous to the paragraph above, i.e., the kernels are plugged in and the summation over $l$ is replaced by an integration over energy $\omega$. Starting from the master equations of $\langle\bf{S^\pm}\rangle$, see Eq. (\ref{eq:MESp}), and by using the kinetic equation of $\langle S_z\rangle$, see Eq. (\ref{eq:MEdiagonal}), one finally obtains the following coordinate-free representation of the master equation for the island spin ${\bf S}$:
	\begin{eqnarray}\label{eq:meqS3parts}
		\dt{\langle{\bf S}\rangle}=\left(\dt{\langle{\bf S}\rangle}\right)_\text{acc}+\left(\dt{\langle{\bf S}\rangle}\right)_\text{rel}+\left(\dt{\langle{\bf S}\rangle}\right)_\text{rot}.
	\end{eqnarray}
	Accumulation, relaxation, and rotation processes determine the time evolution of ${\bf S}$. The contribution, which builds up an average spin, reads
	\begin{eqnarray}
		\nonumber\left(\dt{\langle{\bf S}\rangle}\right)_\text{acc}&=&\frac{\pi\hbar}{2}\sum_{Nr\sigma}p_r\frac{\Gamma_r}{\Gamma_\sigma^r}\left[\hat{{\bf n}}_r+(\hat{{\bf n}}_r\cdot\hat{{\bf n}}_S)\hat{{\bf n}}_S\right]\\
		&&\times\left[\alpha_{r\sigma}^{-}(\Delta_{N-1} ) -\alpha_{r\sigma}^{+}(\Delta_{N} )\right] P_N .
		\label{eq:meSacc}
	\end{eqnarray}
	Transitions to a charge state $N$ by tunneling of electrons from the leads onto the island and vice versa are described by this term and lead to a polarization of the island. On the one hand, tunneling processes of both leads $r$ accumulate a spin in the direction of the respective polarization $\hat{{\bf n}}_r$. But on the other hand, due to the macroscopic spin of the central electrode, there is an additional accumulation contribution in the direction of ${\bf S}$. In equilibrium ($V=0$), there is no current flowing through the central electrode and the equation above becomes $\left(\dt{{\bf S}}\right)_\text{acc}=0$. Hence, without any applied bias voltage, there is no spin accumulated on the island, i.e. ${\bf S}=0$. In contrast to the accumulation contribution, the relaxation term
	\begin{eqnarray}
		\nonumber\left(\dt{\langle{\bf S}\rangle}\right)_\text{rel}\!\!\!\!&=&\!\!\!-\!\sum_{Nr\nu}P_N\!\!\int\!\!\text{d}\omega\;\Gamma_r{\bf s}(\omega)\\
		&&\times\!\!\left[f_{r}^{-}(\omega+\Delta_{N-1} )-f_{r}^{+}(\omega+\Delta_N )\right]\!,
	\end{eqnarray}
	causes a decay of the island spin, which is characterized by ${\bf s}(\omega)$. The vector 
	\begin{equation}
		{\bf s}(\omega)=\frac{\hbar\rho_I}{2}[f_\up(\omega)-f_\dn(\omega)] \hat{{\bf n}}_S
	\end{equation}
	corresponds to the energy-dependent spin density in the central electrode. The influence of the exchange field between ferromagnetic lead $r$ and island is represented by the third contribution of Eq. (\ref{eq:meqS3parts}): 
	\begin{eqnarray}\label{eq:meSrot}
		\left(\dt{\langle{\bf S}\rangle}\right)_\text{rot}\!\!\!\!\!\!&=&\!\!-\frac{g\mu_B}{\hbar}\sum_{r}\int\!\!\text{d}\omega\;{\bf s}(\omega)\times{\bf B}^{r}_\text{exc}(\omega ).
	\end{eqnarray}
	Here, we used the dimensionless magnetic moment of electrons $g$, the Bohr magneton $\mu_B$, and the definition of the exchange field between island and lead $r$:
	\begin{eqnarray}\label{eq:ExcF}
		{\bf B}^{r}_\text{exc}(\omega)&=& \frac{p_r\Gamma_r N_c}{2\pi g\mu_B}\hat{{\bf n}}_r\sum_{N}P_{N} \times\nonumber \\
		&& \int' \text{d}\omega'\!\!\left[\frac{f_r^-(\omega')}{\omega'-\omega-\Delta_N}+\frac{f_r^+(\omega')}{\omega'-\omega-\Delta_{N-1} }\right],
	\end{eqnarray}
	where the prime at the integral denotes a principal value integral. The exchange field can be interpreted as a many-body interaction effect. In the limit of noninteracting island electrons ($E_C=0$) it vanishes independent of gate voltage. The contributions ${\bf B}^{r}_\text{exc}(\omega)$ act on the spin like an applied external magnetic field, which points in the polarization direction of lead $r$. This results in a precession of the accumulated spin out of the ($\hat{{\bf n}}_L$,$\hat{{\bf n}}_R$) plane. Equation (\ref{eq:ExcF}) is similar to the expression of an exchange field existent between a single-level quantum dot and ferromagnetic leads, which was first introduced by Braun \textit{et al.} in the context of a quantum-dot spin valve.\cite{braun:2004} But in contrast to the quantum-dot spin valve, there is a net spin accumulated on the island in all possible charge states $P_{N}$. Hence the total exchange field is a composition of all charge-state contributions. Additionally, the integration over $\omega$ in Eq. (\ref{eq:meSrot}) results from the continuous density of island states.

	Having solved the master equations for $P_N$ and $\langle{\bf S}\rangle$, the stationary charge current through lead $r$ may be calculated. We start from $I_r=\sum_{\chi\chi'}\left(W^{I_r}\right)_{\chi\,\chi'}^{\chi\,\chi'}P_{\chi'}$. Executing the same procedure as in the previous considerations of the master equations one gets
	\begin{eqnarray}
		I_r=e\pi\sum_{N\sigma}\left[\alpha_{r\sigma}^{+}(\Delta_N)-\alpha_{r\sigma}^{-}(\Delta_{N-1})\right]P_{N} \, .
	\end{eqnarray}
	
\section{Results}\label{sec:results}
	We consider electric transport both in the linear and the nonlinear regimes. For this, we calculate the accumulated spin ${\bf S}$ and the (linear and nonlinear) electric current as a function of the system parameters such as gate voltage $V_G$, bias voltage $V$, angle $\phi$ between source and drain's magnetization direction, and the degree of polarization $p$. An explicit focus in the discussion is put on the influence of the exchange field on the transport characteristics.
		
	\subsection{Linear-response regime}
		We start by considering the single-electron spin-valve transistor in the linear-response regime ($eV\ll k_BT,E_C$). All quantities are periodic in the gate voltage $V_G$ with periodicity $e/C_G$. For low temperatures, at most two charge states $N_0$ and $N_0+1$ have a non-vanishing occupation probability. All analytic formulas for the linear-response regime are derived for this limit. With increasing temperature, more charge states may become occupied. The results plotted in the figures are always calculated taking into account all relevant charge states.
		   
\begin{figure}
			\includegraphics[width=.9\columnwidth]{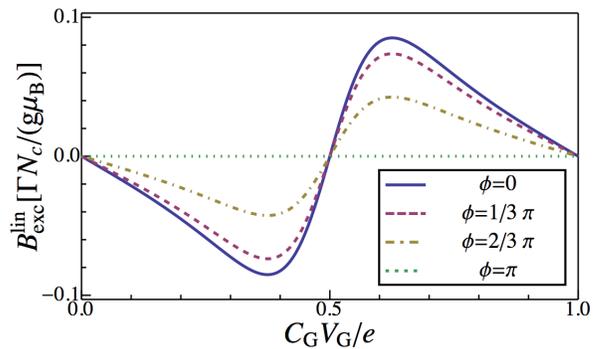}
			\caption{(Color online) Magnitude of exchange field in linear response as a function of applied gate voltage for different angles between the lead polarization directions. Parameters $p_L=p_R=0.3$, $\Gamma_L=\Gamma_R=\Gamma/2$, and $E_C=10k_BT$ were chosen.}
		\label{fig:exclr}
		\end{figure}

		To describe the linear-response regime, the system of master equations Eqs.~(\ref{eq:mefinalP*}) and (\ref{eq:meqS3parts}) is expanded up to first order in the transport voltage $V$. The term associated with rotation of the accumulated island spin simplifies to
		\begin{eqnarray}
			\left(\dt{\langle{\bf S}\rangle}\right)^\text{\text{lin}}_\text{rot}=\frac{g\mu_B}{\hbar}\langle{\bf S}\rangle\times{\bf B}^\text{\text{lin}}_\text{exc}\; ,
		\end{eqnarray}
		where we used the expression of the exchange field in linear response
		\begin{eqnarray}
			{\bf B}_\text{exc}^\text{\text{lin}}=\sum_r\int\!\!\text{d}\omega\;f'(\omega)\left.{\bf B}^{r}_\text{exc}(\omega)\right|_{V=0}\;.
		\end{eqnarray}
		The magnitude $B_\text{exc}^\text{\text{lin}}=|{\bf B}_\text{exc}^\text{\text{lin}}|$ is plotted for different angles $\phi$ as a function of the gate voltage in Fig. \ref{fig:exclr}.
		 The exchange field vanishes at the symmetry point between two resonances (middle of Coulomb valley), i.e., at integer values of $C_GV_G/e$, in accordance to the case of the quantum-dot spin valve. At half-integer values of $C_GV_G/e$, where the two charge states $N_0$ and $N_0+1$ are degenerate $(\Delta_{N_0}=0)$, the exchange field vanishes due to a cancellation of the $N_0$ and $N_0+1$ contributions of ${\bf B}^{r}_\text{exc}(\omega)$. This is in contrast to the case of a quantum-dot spin valve, where the exchange field is maximal at the resonance position for transport. To understand this difference we consider the resonance between charge state 0 (empty dot) and 1 (singly occupied dot) of the single-level quantum dot. Here, the symmetry is broken due to the fact that on the one hand the state 1 can be virtually excited to the doubly-occupied state 2 but on the other hand there is no corresponding virtual excitation of the state 0 in a lower charge state. Hence the different contributions do not cancel each other.
				
\begin{figure}
		 	\includegraphics[width=.7\columnwidth]{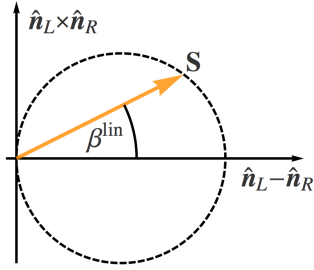}
			\caption{(Color online) Scheme of accumulated spin ${\bf S}$ in linear response regime. The spin precesses within a plane defined by $\hat{{\bf n}}_L-\hat{{\bf n}}_R$ and $\hat{{\bf n}}_L\times\hat{{\bf n}}_R$. The spin precession is accompanied by a decreased magnitude of accumulated spin.}
			\label{fig:SCoordlr}
		\end{figure}

\begin{figure}
		 	\includegraphics[width=\columnwidth]{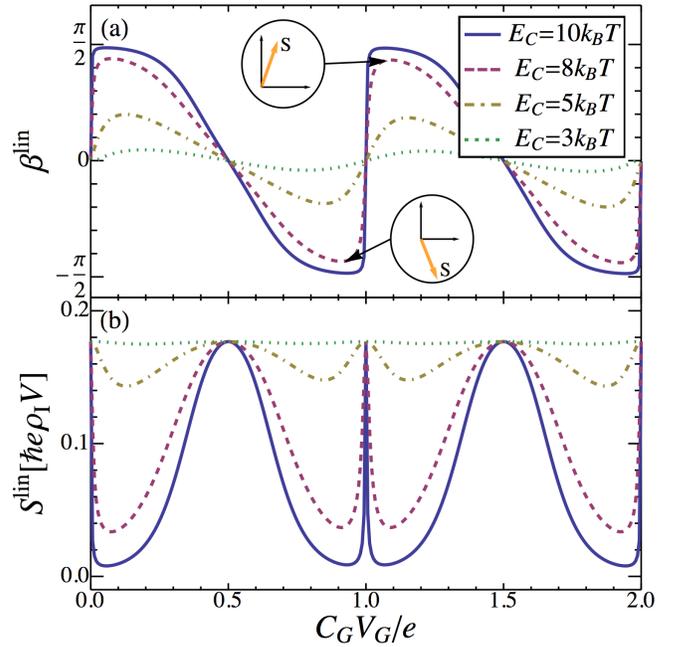}
			\caption{(Color online) (a) Angle $\beta^\text{\text{lin}}$ and (b) spin accumulation $S^\text{\text{lin}}$ for different temperatures as a function of gate voltage. The chosen parameters are $p_L=p_R=0.5$, $\Gamma_L=\Gamma_R=\Gamma/2$, and $\phi=\pi/2$.}
			\label{fig:Slr}
		\end{figure}

		 In the following,  we assume symmetric polarizations ($p_L=p_R=p$) and tunnel-coupling strengths ($\Gamma_L=\Gamma_R=\Gamma/2$). In this case, the total exchange field points in the $\hat{{\bf n}}_L+\hat{{\bf n}}_R$-direction and rotates the spin out of the ($\hat{{\bf n}}_L$,$\hat{{\bf n}}_R$) plane, see Fig. \ref{fig:SCoordlr}. The island spin {\bf S} acquires a component perpendicular to the ($\hat{{\bf n}}_L$,$\hat{{\bf n}}_R$) plane. We obtain $\alpha^\text{\text{lin}}=0$ and
		 \begin{eqnarray}
		 	\beta^\text{\text{lin}}(\phi)=-\arctan\left(\frac{4g\mu_Bk_BT}{\Gamma N_c\Delta_{N_0}}B_\text{exc}^\text{\text{lin}}(\phi)\sinh\frac{\Delta_{N_0}}{k_BT}\right).
		\end{eqnarray}
		In Fig. \ref{fig:Slr}(a), the rotation angle is plotted as a function of the gate voltage. Sign changes appear both at integer and at half-integer values of $C_GV_G/e$, which reflects the sign change of the exchange field at these positions. Away from these points, the angle $\beta^\text{\text{lin}}$ is finite. Close to but not exactly at integer values of $C_GV_G/e$ and at low temperatures ($E_C\approx10k_BT$), the angle approaches $\pm\pi/2$. There, a small variation $V_G$ induces a strong reorientation of the spin {\bf S}, as illustrated by the insets of Fig. \ref{fig:Slr}(a). This high sensitivity can be quantified by the slope of $\beta^\text{\text{lin}}$ with respect to $V_G$. For low temperature ($k_BT\ll E_C$), we find that the slope is proportional to $k_BTp\cos\frac{\phi}{2}\exp{(E_C/k_BT)}/E_C^2$, which increases exponentially with decreasing temperature. As a consequence, the gate-voltage range close to integer values of $C_GV_G/e$ is ideal for manipulating the direction of the island spin. 
		
		The absolute value $S$, which is determined by the accumulation and relaxation term of the spin master equation reads
		 \begin{eqnarray}
		 	S^\text{\text{lin}}(\phi)=\frac{\hbar\rho_Ie}{2}Vp\sin\frac{\phi}{2}\cos{\beta^\text{\text{lin}}(\phi)}.
		 \end{eqnarray}
		 It is proportional to the cosine of $\beta^\text{\text{lin}}$, hence the magnitude decreases with increasing angle $\beta^\text{\text{lin}}$ and vanishes for $\beta^\text{\text{lin}}=\pm\pi/2$, see Fig.~\ref{fig:SCoordlr}. Thus, the effect of the exchange field is not just an rotation of ${\bf S}$ but also a reduction of its magnitude $S$. In Fig.~\ref{fig:Slr}(b), $S^\text{\text{lin}}(\pi/2)$ is plotted for various temperatures. We see that for $E_C\lesssim 3k_BT$ the structure is smeared out and the absolute value of spin is nearly constant due to a suppression of the Coulomb blockade.
		
\begin{figure}
		 	\includegraphics[width=.9\columnwidth]{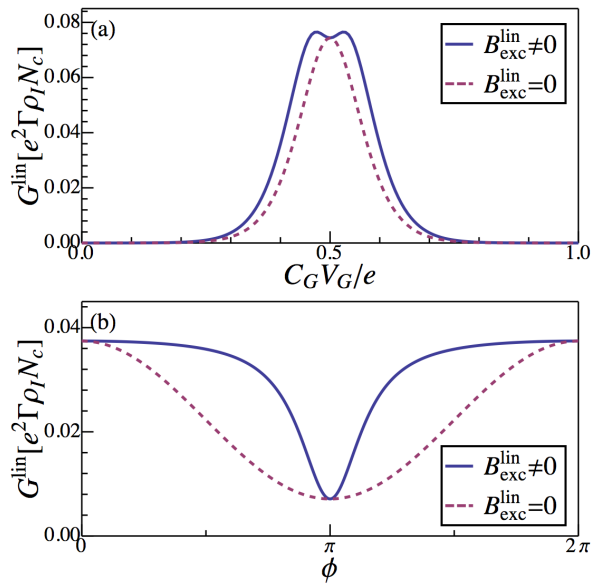}
			\caption{(Color online) Linear conductance in units of $e^2\Gamma\rho_IN_c$ with (solid) and without (dashed) the effect of the exchange field: (a) $V_G$-dependence in case of $\phi=\pi/2$ and (b) $\phi$-dependence with applied gate voltage $C_GV_G=4e/10$. For both plots the parameters $p_L=p_R=0.9$, $\Gamma_L=\Gamma_R=\Gamma/2$, and $E_C=15k_BT$ were chosen.}
			\label{fig:ClrVGPhi}
		\end{figure}

		After considering the accumulated spin, we now analyze the linear conductance $G^\text{\text{lin}}=(\partial I/\partial V)|_{V=0}$ of the single-electron spin-valve transistor. The conductance, normalized to the conductance of a single-electron spin-valve transistor with parallel lead magnetizations $G^\text{\text{lin}}(0)=e^2N_c\rho_I\Gamma\Delta_{N_0}/(8 k_BT \sinh\frac{\Delta_{N_0}}{k_BT})$, can be expressed in terms of the rotation angle $\beta^\text{\text{lin}}(\phi)$ as
		\begin{eqnarray}\label{eq:condlr}
			\frac{G^\text{\text{lin}}(\phi)}{G^\text{\text{lin}}(0)}=1-\frac{p^2\sin^2\frac{\phi}{2}}{1+\tan^2\beta^\text{\text{lin}}(\phi)}\;.
		\end{eqnarray}
		In Fig. \ref{fig:ClrVGPhi}(a), the conductance $G^\text{\text{lin}}(\phi=\pi/2)$ is plotted as a function of gate voltage. Since it is an even function of charging energy $\Delta_{N_0}$, the conductance is symmetric with respect to the resonance points, given by half-integer values of $C_GV_G/e$. Away from the resonances, the conductance is suppressed due to Coulomb blockade. The difference between the solid and dashed lines illustrates the influence of the exchange field. The dashed lines are obtained by manually setting $B_\text{exc}^\text{\text{lin}}$ to zero. The exchange field rotates the accumulated spin out of its blocking position. Hence, it increases the conductance of the single-electron spin-valve transistor except for the symmetry point at resonance. As a consequence, for high polarizations, $p\gtrsim0.9$, the maximum at resonance can even turn into a local minimum.  
		
		Consideration of the conductance as a function of the magnetization angle $\phi$, see Fig. \ref{fig:ClrVGPhi}(b), shows that the exchange field reduces the spin-valve effect. We emphasize that, though the solid and dashed lines coincide for $\phi\in\{0,\pi,2\pi\}$, the exchange field does not vanish in these points. In fact, it points parallel to the accumulated spin and, therefore, ${\bf B}_\text{exc}^\text{\text{lin}}$ does not rotate ${\bf S}$ out of the ($\hat{{\bf n}}_L$,$\hat{{\bf n}}_R$)-plane.

	\subsection{Nonlinear-response regime}
\begin{figure}
		 	\includegraphics[width=.9\columnwidth]{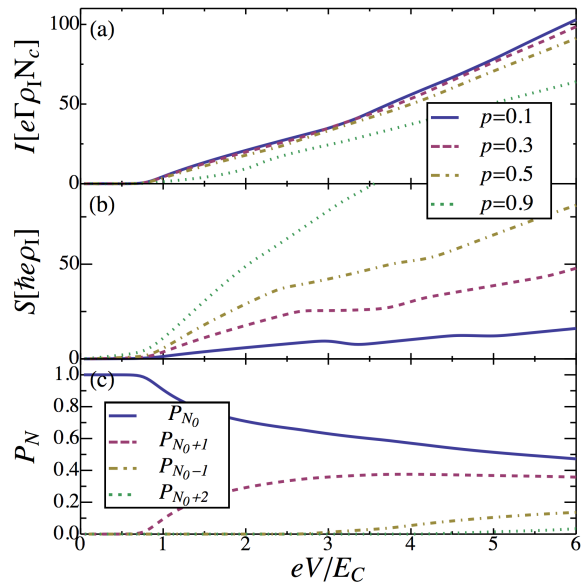}
			\caption{(Color online) (a) Current $I$ through single-electron spin-valve transistor in units of $e\Gamma\rho_IN_c$ for different polarizations $p$, (b) the corresponding island spin accumulation $S$, and (c) the relevant occupation probabilities $P_{N}$ in the case $p=0.3$. For all three plots, the parameters $p_L=p_R=p$, $\Gamma_L=\Gamma_R=\Gamma/2$, $\phi=\pi/2$, $C_GV_G=3e/10$, and $E_C=50k_BT$ were chosen.}
			\label{fig:IVnlrMulti}
		\end{figure}

		We now turn to the nonlinear-response regime, $eV\gtrsim k_BT,E_C$. The transport voltage $V$ is symmetrically applied to both leads, i.e., the chemical potentials are $\mu_L=eV/2$ and $\mu_R=-eV/2$. The current through the single-electron spin-valve transistor as a function of bias voltage for different polarizations $p$ is shown in Fig. \ref{fig:IVnlrMulti}(a). Here, the angle between the lead magnetizations is chosen to be $\phi=\pi/2$. Due to the larger spin accumulation for higher lead polarizations, the current reduces with increasing $p$. At low bias voltages, the transport through the system is blocked until $eV$ exceeds the Coulomb blockade threshold $2\Delta_{N_0}$ and both spin reservoirs of the next charging state ($N_0+1$) enter the transport window. We introduce the notation $E_{N;\sigma}$ for the excitation energy of the $\sigma$ channel of charge state $N$. From this definition it follows that $E_{N_0+1;\sigma}=2\Delta_{N_0}$. By further increasing of the source-drain voltage, more and more levels of the continuous spectrum contribute to transport, and the current increases continuously (in contrast to a stepwise increase for quantum dot with a discrete level spectrum). We note that we assumed here symmetric tunnel couplings, $\Gamma_L=\Gamma_R=\Gamma/2$. By an asymmetric choice of the tunnel couplings, the Coulomb step at $eV=2\Delta_{N_0}$ and also the steps of the other charging states that enter for higher $V$ can be made much more pronounced.\cite{amman:1991,averin:1991} In Fig. \ref{fig:IVnlrMulti}(c), the corresponding occupation probabilities of the relevant charging states ($N\in\{N_0-2,N_0-1,N_0,N_0+1\}$) are plotted as a function of the bias voltage (for $p=0.3$). The probability to find the island in a given charge state $N$ decreases for states with higher excitation energies. In general, the voltage that is necessary to excite the island state ($N,\sigma$) is determined by the equation $eV=|2\Delta_N+\sigma\Delta\mu(V)|$. 	Here, $\Delta\mu(V)$ is the island level splitting for the respective bias voltage and $\sigma$ contributes with a factor +1 for the $\up$ reservoir and -1 for the $\dn$ reservoir. It is proportional to the spin accumulation $S(\phi)$, which is plotted in Fig. \ref{fig:IVnlrMulti} (b). By means of $\Delta\mu(V)$ the transition voltages can be determined self consistently. We emphasize that due to a finite island spin accumulation, either the spin $\up$- or spin $\dn$ reservoir enters the transport window at the transition voltages of the occupation probabilities, see Fig. \ref{fig:energy}.
		
		 An analysis of the bias-voltage dependence of the island spin [see Fig. \ref{fig:IVnlrMulti} (b)] yields that $S$ can be decomposed into two components, an oscillating one and a component that monotonously increases.\cite{ernult:2007,barnas:2008} Since the former is suppressed for higher $p$, let us consider the graph of $p=0.1$ to explain this behavior. Our theory describes sequential tunneling, hence in the Coulomb-blockade regime, spin accumulation is exponentially suppressed. For $eV>2\Delta_{N_0}$, both spin reservoirs of the charging state $N_0+1$ contribute to transport and $S$ increases with increasing $V$ until the $\up$ reservoir of charging state $N_0-1$ enters the transport window. Now, there is an additional channel for the $\up$ electrons to leave the island and $S$ decreases until the corresponding $\dn$ reservoir is reached. Subsequently, the spin accumulation increases until $\dn$ electrons can tunnel into charging state $N_0+2$. A reduction of $S$ follows, then the state $\up,N_0+2$ can be occupied and the spin accumulation increases again. By further increasing of $V$ this processes of depletion and filling repeat. To understand the reduced influence of the oscillating component for higher $p$, we first assume antiparallely polarized leads $\phi=\pi$. In this case, the tunneling rates of the processes, which lead to a decrease of $S$, are proportional to $1-p$, hence they occur less for higher $p$. Since there is a more complex but similar behavior of the tunneling rates for finite noncollinear angles $\phi$, we can conclude that, in general, the oscillations are weaker for higher polarizations.
		
\begin{figure}
		 	\includegraphics[width=.9\columnwidth]{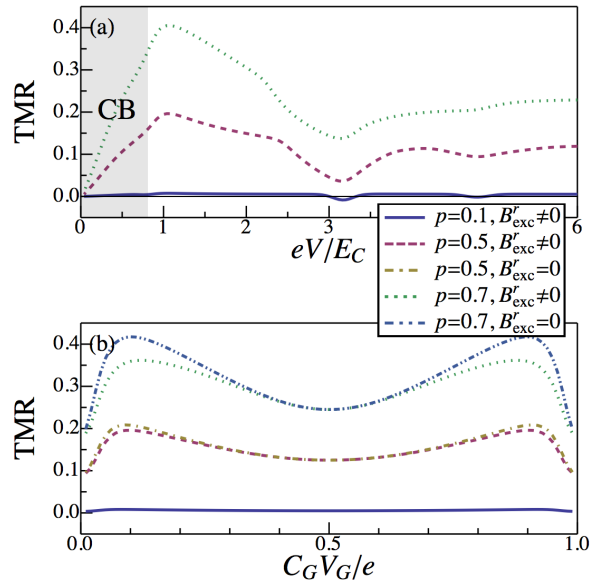}
			\caption{(Color online) TMR of a noncollinear setup ($\phi=\pi/2$) for different polarizations $p$. Plot (a) shows the bias dependence for $C_GV_G=3e/10$ and (b) the gate dependence for $eV=2E_C$. The dashed-dotted lines represent the cases where the exchange field is manually set to ${\bf B}_\text{exc}^{r}=0$. For both plots, the remaining parameters were chosen to be $p_L=p_R=p$, $\Gamma_L=\Gamma_R=\Gamma/2$, and $E_C=50k_BT$.}
			\label{fig:TMRnlr}
		\end{figure}

		The finite spin accumulation on the island results in a nonzero TMR. Due to the noncollinear system we define an angular-dependent magnetoresistance:
		\begin{eqnarray}
			\text{TMR}(\phi)=\frac{I(0)-I(\phi)}{I(0)},
		\end{eqnarray}
		with $I(\phi)$ being the current through the single-electron spin-valve transistor in which the magnetization directions of the leads enclose the angle $\phi$. In Fig. \ref{fig:TMRnlr}, the bias- and gate-voltage dependence of the TMR is plotted for different polarizations and an angle $\phi=\pi/2$. In both cases, the TMR shows an oscillatory behavior originating from the bias/gate dependence of the spin accumulation.\cite{majumdar:1998,korotkov:1999,brataas:1999a,brataas:1999b,weymann:2003} Naturally, the magnetoresistance is more pronounced for higher $p$. For lower polarizations, we observe sign changes of TMR caused by the oscillating spin accumulation.\cite{ernult:2007} We want to emphasize that for obtaining reliable results in the Coulomb-blockade regime [marked by the grey area in Fig. \ref{fig:TMRnlr}(a)], one has to take cotunneling processes into account, which we neglect in our theory. Next, we focus on the analysis of the influence of the exchange field on the TMR, which is illustrated by the dashed-dotted lines in Fig. \ref{fig:TMRnlr}(b). To obtain the lines representing the case of an absent exchange field, we manually set ${\bf B}_\text{exc}^{r}=0$ in Eq.~(\ref{eq:meqS3parts}). For gate voltages representing a vanishing exchange field ($C_GV_G/e\in\{0,1/2,1\}$), naturally, the two graphs coincide. Comparison of the TMR gate dependence for different polarizations shows, that the exchange field has a stronger effect for higher $p$. By affecting the accumulated spin the exchange field decreases the TMR. 
		
\begin{figure}
		 	\includegraphics[width=.9\columnwidth]{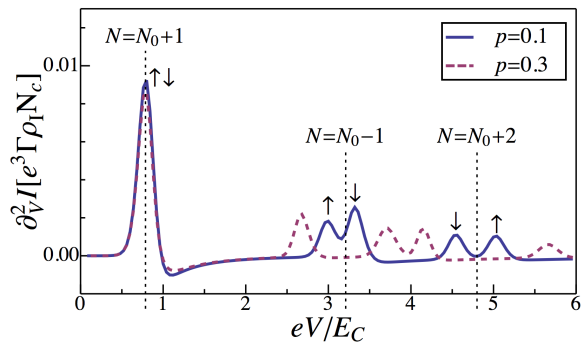}
			\caption{(Color online) Second derivative of the current $\partial^2 I/\partial V^2$ in units of $e^3\Gamma\rho_IN_c$ for two different polarizations $p$ and $\phi=\pi/2$. The excitation energies of the charge/spin states represented by the peaks strongly depend on the lead polarization $p$. Parameters $p_L=p_R=p$, $\Gamma_L=\Gamma_R=\Gamma/2$, $C_GV_G=3e/10$, and $E_C=50k_BT$ were chosen.}
			\label{fig:CVnlr}
		\end{figure}

\begin{figure}
		 	\includegraphics[width=.9\columnwidth]{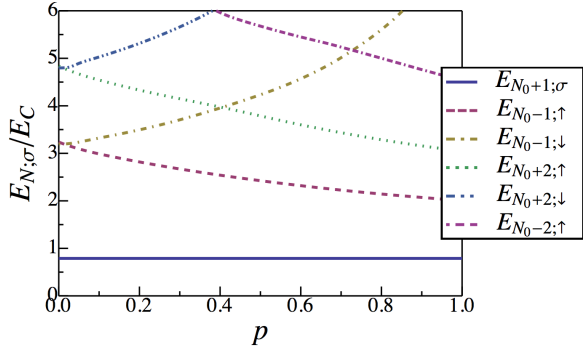}
			\caption{(Color online) Polarization dependence of the excitation energies $E_{N;\sigma}$ normalized to $E_C$. The chosen parameters are $p_L=p_R=p$, $\phi=\pi/2$, $\Gamma_L=\Gamma_R=\Gamma/2$, $C_GV_G=3e/10$, and $E_C=50k_BT$.}
			\label{fig:Cpnlr}
		\end{figure}

		In order to identify the threshold voltages at which new transport channels open, it is convenient to study the second derivative of the current $\partial^2 I/\partial V^2$, see Fig. \ref{fig:CVnlr}.
The positions of the peaks directly represent the already discussed excitation energies of the relevant charging states. All charging states induce two peaks, one for each spin channel ($\up,\dn$). An exception is the first peak  ($N_0+1$), which does not split into two due to the exponentially suppressed spin accumulation at the relevant voltage, see Fig. \ref{fig:IVnlrMulti}(b). The vertical lines in the figure mark the voltages at which the charge states ($N\in\{N_0-1,N_0+1,N_0+2\}$) would enter the transport window if one neglects spin accumulation. Due to the fact that the accumulated island spin strongly depends on the lead polarizations, the excitation energies $E_{N;\sigma}$ are also sensitive to a variation of $p$, see Fig. \ref{fig:Cpnlr}. The described behavior suggests that in experiments, a measurement of $\partial^2 I/\partial V^2$ as a function of bias voltage can be used as a convenient tool to determine the spin splitting of the chemical potential on the island and, thus, the degree of polarization of the leads.

\section{Conclusion}    \label{sec:conclusion}
	We have investigated electronic transport through a noncollinear single-electron spin-valve transistor realized by a metallic island weakly coupled to two ferromagnetic leads. Employing a diagrammatic real-time technique, we performed a systematic perturbation expansion in the tunnel-coupling strength. We found that an interaction-induced exchange field appears as a consequence of a spin-dependent tunnel coupling to the leads. In the linear-response regime, this exchange field weakens the spin-valve effect and increases the linear conductance. It, furthermore, leads to a high sensitivity of orientation of the accumulated island spin on variation of the gate voltage $V_G$, which is convenient for controlled manipulation of the island spin. Finally, we suggest the second derivative of the current $\partial^2 I/\partial V^2$ in the nonlinear regime as a tool to determine the spin splitting of the chemical potential on the island, which, in turn, depends on the degree of spin polarization in the leads of the single-electron spin-valve transistor. 

\acknowledgments
	We acknowledge financial support from DFG via grant KO 1987/4. 

\appendix

\section{Kernel elements of master equations} \label{ap:kerele}
	In the calculation of the matrix elements of the kernel $\bf W$ to first order in the tunnel coupling only one island level $l$ and channel number $\nu$ is involved. Its value depends on the energy and the occupation of this level as well as on the total island charge. As discussed in the main text, we need matrix elements that describe transitions from a state that is diagonal in the occupation of the considered island level to a state that may be diagonal (for the kinetic equations of $P_N$ and $S_z$) or off-diagonal (for the kinetic equations of $S^\pm$). Explicit calculation yields
	\begin{eqnarray}
 		W_{\sigma \, 0}^{\sigma \, 0}(l,\nu,N) &=&\frac{1}{\hbar}\sum_{rs}\Gamma^r_{s\sigma}f^+_r(\epsilon_l+\Delta_{N})\, ,\\
		W_{0 \, \sigma}^{0 \, \sigma}(l,\nu,N) &=&\frac{1}{\hbar}\sum_{rs}\Gamma^r_{s\sigma}f^-_r(\epsilon_l+\Delta_{N-1})\, ,\\
 		W_{d \, \sigma}^{d \, \sigma}(l,\nu,N) &=&\frac{1}{\hbar}\sum_{rs}\Gamma^r_{s\bar\sigma}f^+_r(\epsilon_l+\Delta_{N})\, ,\\
		W_{\sigma \, d}^{\sigma \, d}(l,\nu,N) &=&\frac{1}{\hbar}\sum_{rs}\Gamma^r_{s\bar\sigma}f^-_r(\epsilon_l+\Delta_{N-1})\, ,
	\end{eqnarray}
	where $f_r^+(E)=f_r(E)$ describes tunneling of electrons from lead $r$ into the island and $f_r^-(E)=1-f_r(E)$ tunneling out of the island into lead $r$. Hence the elements above describe processes that change the occupation of the central electrode. These parts of the kernel are related to the diagonal matrix elements where the total island charge remains constant via
	\begin{eqnarray}
		W_{0 \, 0}^{0 \, 0}(l,\nu,N) &=&\!\!\!- \sum_\sigma W_{\sigma \, 0}^{\sigma \, 0}(l,\nu,N),\\
		W_{\sigma \, \sigma}^{\sigma \, \sigma}(l,\nu,N) &=&\!\!\! - W_{0 \, \sigma}^{0 \, \sigma}(l,\nu,N) - W_{d \, \sigma}^{d \, \sigma}(l,\nu,N),\\
		W_{d \, d}^{d \, d}(l,\nu,N) &=&\!\!\! - \sum_\sigma W_{\sigma \, d}^{\sigma \, d}(l,\nu,N).
	\end{eqnarray}
	The matrix elements with off-diagonal final states can also be divided into those where the total island charge changes during the transition, 
	\begin{eqnarray}
 		\!\!\!\!\!\!W_{\bar\sigma \, 0}^{\sigma \, 0}(l,\nu,N)\!\!\!&=&\!\!\!\frac{2\pi}{\hbar}\sum_{rs}\rho_s^rV^{r^*}_{s\sigma}V^r_{s\bar{\sigma}}f_r^+(\epsilon_l+\Delta_N ),\\
	 	\!\!\!\!\!\!W_{\bar\sigma \, d}^{\sigma \, d}(l,\nu,N)\!\!\!&=&\!\!\!-\frac{2\pi}{\hbar}\sum_{rs}\rho_s^rV^{r^*}_{s\sigma}V^r_{s\bar{\sigma}}f_r^-(\epsilon_l+\Delta_{N-1} ), 
	\end{eqnarray}
	and those where the total island charge remains constant,
	\begin{eqnarray}
 		W_{\bar\sigma \, \up}^{\sigma \, \up}(l,\nu,N)&=&-\frac{2\pi i}{\hbar}\sum_{rs}\rho_s^rV^{r^*}_{s\sigma}V^r_{s\bar{\sigma}}I(N),\\
		W_{\bar\sigma \, \dn}^{\sigma \, \dn}(l,\nu,N)&=&\frac{2\pi i}{\hbar}\sum_{rs}\rho_s^rV^{r^*}_{s\sigma}V^r_{s\bar{\sigma}}I^{^*}(N)\, ,	
	\end{eqnarray}
	with the integral expression 
	\begin{eqnarray}
		\nonumber I(N)=\int\frac{\text{d}\omega}{2\pi}\left[\frac{f^-_r(\omega)}{\epsilon_l+\Delta_{N-1}-\omega+i0^+}\right.\\
		\left.+\frac{f^+_r(\omega)}{\epsilon_l+\Delta_{N}-\omega-i0^+}\right],
	\end{eqnarray}
	which represents transition processes of island-charge states, while $\operatorname{Re}I(N)$ exclusively describes virtual charge transfer.

\end{document}